\documentclass[fleqn,twoside]{article}
\usepackage{espcrc2}

\usepackage{graphicx}
\usepackage[figuresright]{rotating}

\newcommand{\AmS}{{\protect\the\textfont2
  A\kern-.1667em\lower.5ex\hbox{M}\kern-.125emS}}

\makeatletter
\def\simleq{\mathrel{\mathpalette\gl@align<}}
\def\simgeq{\mathrel{\mathpalette\gl@align>}}
\def\gl@align#1#2{\lower.6ex\vbox{\baselineskip\z@skip\lineskip\z@
     \ialign{$\m@th#1\hfill##\hfil$\crcr#2\crcr\sim\crcr}}}
\makeatother

\newcommand{\bra}{\langle}
\newcommand{\ket}{\rangle}
\newcommand{\braket}[1]{\bra #1 \ket}
\newcommand{\qq}{\braket{\bar{q}q}}

\newcommand{\qGq}{g\braket{\bar{q}\sigma_{\mu\nu}G_{\mu\nu} q}}
\newcommand{\nn}{\nonumber\\}

\title{
Thermal Effects on the Condensates of
Chiral Order Parameters,\\
$\langle\bar{q}q\rangle$
and
$g\langle\bar{q}\sigma_{\mu\nu}G_{\mu\nu} q\rangle$,
and Chiral Restoration from Lattice QCD
}

\author{
Takumi Doi%
\address[titech]{
Department of Physics, Tokyo Institute of Technology,\\
Ohokayama 2-12-1, Meguro, Tokyo 152-8551, Japan}, 
Noriyoshi Ishii%
\address[riken]{
Radiation Laboratory, 
The Institute of Physical and Chemical Research (RIKEN),\\
Hirosawa 2-1, Wako, Saitama, 351-0198, Japan},
Makoto Oka\addressmark[titech]
and
Hideo Suganuma\addressmark[titech]
}

\begin{document}

\begin{abstract}

We study the quark-gluon mixed condensate 
$g\langle\bar{q}\sigma_{\mu\nu}G_{\mu\nu} q\rangle$ 
using the SU(3)$_c$ lattice QCD with the Kogut-Susskind fermion 
at the quenched level.
We perform the first but
accurate measurement of the thermal effects 
on the mixed condensate in lattice QCD at  
$\beta=6.0$ with $16^3\times N_t (N_t=16,12,10,8,6,4)$,
$\beta = 6.1$ with $20^3\times N_t (N_t=20,12,10,8,6)$, and 
$\beta = 6.2$ with $24^3\times N_t (N_t=24,16,12,10,8)$.
At each temperature, 
the lattice calculation is carried out with high statistics using more 
than 100 gauge configurations.
In particular, about 1000 gauge configurations are used near the critical 
temperature $T_c$.
The chiral extrapolation is done from the lattice result 
with the current-quark mass of $m_q = 21,36,52 {\rm MeV}$.
While the thermal effects on both of $\qGq$ and $\qq$ 
are very weak except for the vicinity of $T_c$, 
both the condensates suddenly vanish around $T_c \simeq 280 {\rm MeV}$,
which indicates strong chiral restoration near $T_c$.
To compare the thermal effects on the two condensates, 
we analyze the ratio 
$m_0^2 \equiv g\langle\bar{q}\sigma_{\mu\nu}G_{\mu\nu} q\rangle / 
\langle\bar{q}q\rangle$,
and find that $m_0^2$ is almost independent of the temperature,
even in the very vicinity of $T_c$. 
This means that these two different condensates obey 
the same critical behavior,  
and this nontrivial similarity between them would impose constraints on 
the chiral structure of the QCD vacuum near $T_c$.
\vspace{1pc}
\end{abstract}

\maketitle

\section{Introduction}

The non-perturbative nature of QCD provides
various phenomena
such as spontaneous chiral-symmetry breaking and
color confinement.
Currently, the RHIC experiments are in progress
to understand these phenomena, and 
the finite-temperature QCD becomes important.
In this point, 
condensates are useful physical quantities because 
they characterize the nontrivial QCD vacuum.
At finite temperature, 
the QCD vacuum is expected to change, and 
we can examine it from 
the thermal effects on the condensates.
%
%

Among various condensates, we study 
the quark-gluon mixed condensate $\qGq$ for the following reasons.
First, we note that the mixed condensate is 
a chiral order parameter independent of $\qq$,
since it flips the chirality of the quark as
\begin{eqnarray}
\lefteqn{\qGq} \nn
&=&
  g\braket{\bar{q}_R (\sigma_{\mu\nu}G_{\mu\nu}) q_L}
+ g\braket{\bar{q}_L (\sigma_{\mu\nu}G_{\mu\nu}) q_R}.
\end{eqnarray}
Therefore, $\qGq$, as well as $\qq$, is an important indicator 
on the chiral structure of the QCD vacuum.
In particular, 
the thermal effects 
on both the condensates and the comparison of them
will reveal mechanism of the
chiral restoration of the QCD vacuum at finite temperature.
Second, in contrast to $\qq$, 
$\qGq$ represents a direct correlation 
between quarks and gluons in the QCD vacuum, and thus
characterizes different aspect of the QCD vacuum. 
%
%
Third,
the mixed condensate is relevant in 
various QCD sum rules, especially in baryons~\cite{Dosch},
light-heavy mesons~\cite{Dosch2}
and exotic mesons~\cite{Latorre}.
In this point, the quantitative estimate of the thermal
effects on $\qGq$ has an impact on hadron phenomenology at finite temperature
through the framework of the QCD sum rule.
Therefore, it is desirable to estimate $\qGq$ at zero/finite temperature
by a direct calculation from QCD, 
such as lattice QCD simulation.

So far, the mixed condensate $\qGq$ at zero temperature has been 
analyzed phenomenologically in the QCD sum rules~\cite{Bel}.
In the lattice QCD, a pioneering work~\cite{K&S} has been done 
long time ago, but the result was rather preliminary.
Recently, new lattice calculations have been performed 
by us~\cite{DOIS:qGq,DOIS:T} using the Kogut-Susskind (KS) fermion,
and by other group~\cite{twc:qGq} using the Domain-Wall fermion.
At finite temperature, however, there has been no 
result on $\qGq$ except for 
our early results in Ref.~\cite{DOIS:T}.
Therefore, we here present the extensive results
of the thermal effects on $\qGq$ as well as $\qq$,   
including the analysis near the critical temperature.

\section{Lattice formalism and the parameter set}

We calculate the condensates $\qq$ and $\qGq$ 
using the SU(3)$_c$ lattice QCD with the KS-fermion 
at the quenched level.
We note that the KS-fermion preserves the explicit chiral symmetry
for the quark mass $m=0$, which is essential for our study.

The Monte Carlo simulations are performed with the 
standard Wilson action for $\beta=6.0, 6.1$ and $6.2$.
In order to calculate at various temperatures, 
we use the following lattices as
\begin{tabular}{cl}
i)   & $\beta = 6.0$,\  $16^3\times N_t\ (N_t=16,12,10,8,6,4)$,\\
ii)  & $\beta = 6.1$,\  $20^3\times N_t\ (N_t=20,12,10,8,6)$, \\
iii) & $\beta = 6.2$,\  $24^3\times N_t\ (N_t=24,16,12,10,8)$.
\end{tabular}
We generate 100 gauge configurations for each lattice.
However, in the vicinity of the chiral phase transition point,
namely, $20^3\times 8$ at $\beta=6.1$ and $24^3\times 10$ at
$\beta=6.2$, the fluctuations of the condensates get larger.
We hence generate 1000 gauge configurations for the 
above two lattices to improve the estimate.
The lattice units are obtained as
$a\simeq 0.10, 0.09, 0.07 {\rm fm}$ for 
$\beta = 6.0, 6.1, 6.2$, respectively, 
which reproduce 
the string tension $\sigma = 0.89 {\rm GeV/fm}$.
We calculate the flavor-averaged condensates as
\begin{eqnarray}
a^3 \qq
= - \frac{1}{4}\sum_f {\rm Tr}\left[ \braket{q^f(x) \bar{q}^f(x)} \right],
\end{eqnarray}
\begin{eqnarray}
\lefteqn{a^5 \qGq} \nn
&=& - \frac{1}{4}\sum_{f,\ \mu,\nu}{\rm Tr}
        \left[ \braket{q^f(x) \bar{q}^f(x)} \sigma_{\mu\nu} G_{\mu\nu}
\right],
\end{eqnarray}
where SU(4)$_f$ quark-spinor fields, $q$ and $\bar{q}$, are converted into 
spinless Grassmann KS-fields $\chi$, $\bar{\chi}$
and the gauge-link variable.
We adopt the clover-type definition for
the gluon field strength $G_{\mu\nu}$ on the lattice 
in order to eliminate ${\cal O}(a)$ discretization error.
The more detailed formula for the condensates and gluon field strength
are given in Ref.~\cite{DOIS:qGq}.

In the calculation of the condensates,
we use the current-quark mass $m = 21,36,52 {\rm MeV}$.
For the fields $\chi$, $\bar{\chi}$, the anti-periodic condition 
is imposed at the boundary in the all directions.
We measure the condensates 
on 16 different physical space-time points of $x$ ($\beta=6.0$) or 
2 points of $x$ ($\beta = 6.1,6.2$)
in each configuration.
Therefore, at each $m$ and temperature, 
we achieve high statistics in 
total as
1600 data ($\beta=6.0$) or 200 data ($\beta=6.1,6.2$).
Note that in the vicinity of the critical temperature, $20^3\times 8$ 
($\beta =6.1$)
and $24^3\times 10$ ($\beta =6.2$), 
we take up to 2000 data so as to achieve  high statistics 
and to guarantee reliability of the results.


\section{The Lattice results and Discussions}

We calculate the condensates $\qq$ and $\qGq$ 
at each quark mass and temperature.
We observe that both the condensates show 
a clear linear behavior against the quark mass $m$.
Typical plot can be seen in Ref.~\cite{DOIS:qGq}.
Therefore, we fit the data with a linear function and determine 
the condensates in the chiral limit.
Estimated statistical errors are typically less than 5\%,
while the finite volume artifact is estimated to be about 1\%, 
from the dependence of the condensates on
boundary conditions~\cite{DOIS:qGq}.

\begin{figure}[htb]
\begin{center}
\includegraphics[scale=0.28]
{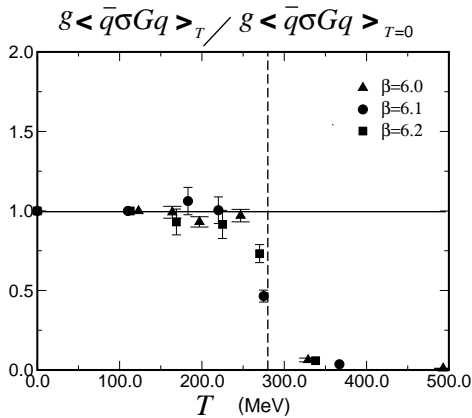}
\end{center}
\vspace*{-10mm}
\caption{
Thermal effects on $\qGq$
plotted against the temperature $T$.
The vertical dashed line denotes the critical 
temperature $T_c \simeq 280 {\rm MeV}$
at the quenched level.
}
\label{fig:qGq_finite_T}
\end{figure}

For each condensate, 
We estimate the thermal effects 
by taking the ratio between the values at finite 
and zero temperature.
This procedure eliminates
the renormalization constants
because there is no operator mixing for 
both the condensates
in the chiral limit~\cite{Narison2}.
In figure~\ref{fig:qGq_finite_T}, we plot the thermal effects on 
$\qGq$.
We find a drastic change of $\qGq$ around the critical temperature
$T_c \simeq 280 {\rm MeV}$. This is a first observation of 
chiral-symmetry restoration  in $\qGq$.
We also find that 
the thermal effects on $\qGq$ are very weak
below the critical temperature, 
namely, $T \simleq 250 {\rm MeV}$.
The same nontrivial features are also found for $\qq$.

\begin{figure}[htb]
\vspace*{-4mm}
\includegraphics[scale=0.28]
{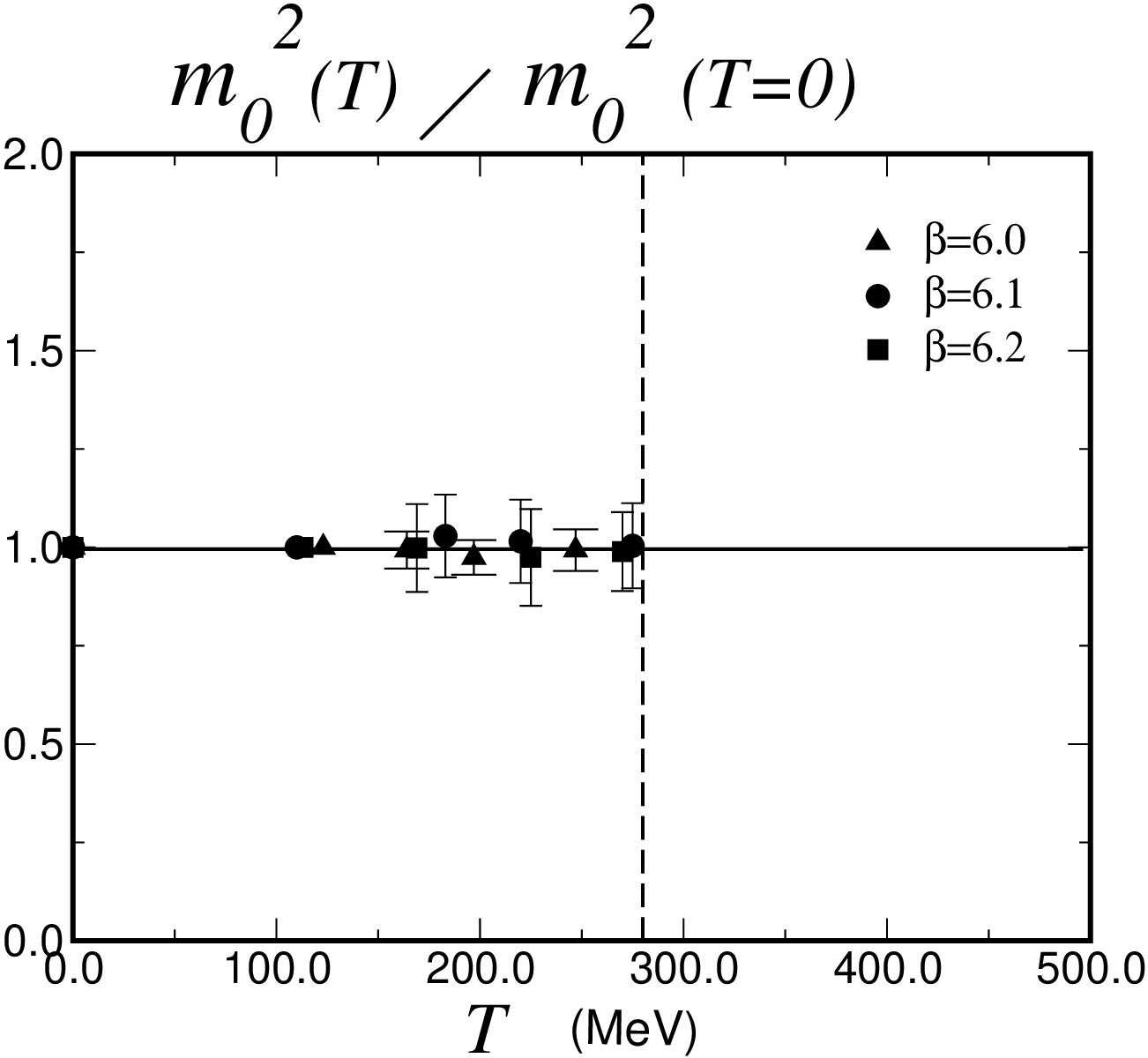}
\vspace*{-3mm}
\caption{The thermal effects on $m_0^2 \equiv \qGq/\qq$ plotted against 
the temperature $T$.
This result indicates the same critical behavior 
between $\qq$ and $\qGq$.
}
\label{fig:M0_finite_T}
\vspace*{-5mm}
\end{figure}

We then compare the thermal effects on
$\qGq$ and $\qq$,
because these two condensates 
characterize different aspects of the QCD vacuum.
For this purpose, we plot the thermal effects 
on $m_0^2 \equiv \qGq / \qq$ in figure~\ref{fig:M0_finite_T}.
%
%
%
%
From figure~\ref{fig:M0_finite_T}, we observe that $m_0^2$ is almost 
independent of the temperature, even in the very vicinity of $T_c$.
This very nontrivial result means that both of $\qGq$ and $\qq$ 
obey the same critical behavior.
This might indicate the existence of the universal 
behavior of chiral order parameters near $T_c$,
and would impose constraints on the chiral structure of the QCD vacuum.

In summary, we have studied the thermal effects on $\qGq$ as well 
as $\qq$ 
from lattice QCD at the quenched level. 
We have found a clear signal of chiral restoration
from both the condensates, while 
the thermal effects below $T_c$ are very weak.
We have also observed that 
$m_0^2 \equiv \qGq /\qq$ 
is almost independent of the temperature, which means
both the condensates obey the same critical behavior.
This nontrivial similarity would impose constraints on 
the chiral structure of the QCD vacuum near $T_c$.
For further studies,
a full QCD lattice calculation is in progress
in order to analyze the dynamical quark effects on the condensates.

\end{document}